\documentclass[pre,amsmath,amssymb]{revtex4}

\usepackage{amssymb}
\usepackage{graphicx}
\usepackage{dcolumn}
\usepackage{bm}
\usepackage[latin1]{inputenc}   
\usepackage[T1]{fontenc}        

\begin{document}




\title{Nucleation and growth of second phase precipitates under
    non-isothermal conditions \footnote{This is an expanded version of
the paper presented in the International Conference on Solid-Solid
Phase Transformation in Inorganic Materials 2005 (PTM 2005), May 29 -
June 3, 2005, Phoenix, Arizona, USA.}}

\author{A. R. Massih$^{1,2}$ and L. O. Jernkvist$^1$}
\affiliation{$^1$ Quantum Technologies, Uppsala Science Park, SE 751 83
  Uppsala, Sweden
}%
\affiliation{%
  $^{2}$Materials Science, Malm\"{o} University, SE 205 06 Malm\"{o}, Sweden
}%

\date{\today}

\begin{abstract}
  The Langer-Schwartz equations for precipitation are formulated to
  calculate nucleation, growth and coarsening of second phase
  precipitates under non-isothermal situations. A field-theoretic
  steady-state nucleation rate model is used in the analysis. The
  field-theoretic nucleation rate is compared with the classical
  nucleation rate relation derived from the Fokker-Planck equation.
  This integrated model is used to simulate a bcc-to-hcp phase
  quenching and subsequent annealing in the hcp phase of a dilute
  zirconium alloy, where the mean precipitate size, number density and the
  degree of supersaturation are calculated as a function of time.  The
  influence of cooling rate on the aforementioned parameters is
  evaluated. A lower cooling rate results in larger precipitates with
  a smaller number density in concordance with observations.
\end{abstract}

\maketitle

\section{Introduction}
\label{sec:Intro}

Various important properties of engineering alloys, such as their
mechanical strength and toughness, creep and corrosion resistance,
magnetic and superconducting behaviors, are basically governed by the
presence and characteristics of second phase precipitated particles
(SPPs) \cite{Wagner_Kampmann_1991}. In many material processing
methods, the SPP characteristic is controlled by quenching the
material from a high temperature phase, where the constituents of the
SPPs are in solid solution, to a low temperature phase, in which
precipitation occurs and a distribution of SPPs is observed. This
process embraces such phenomena as nucleation, spinodal decomposition,
late stage growth and coarsening, which are broadly described by the
framework of the kinetics of first-order transitions
\cite{Gunton_et_al_1983}.

For example, when a binary alloy system is quenched from an
equilibrium state (phase) to a non-equilibrium state inside a
coexistence curve of its phase diagram, the quenched system gradually
transforms from the non-equilibrium state to an equilibrium state,
consisting of two coexisting phases. This transformation occurs by
time-dependent spatial fluctuations of the local concentration of one
of the two component species. Commonly, two different kinds of
instabilities are considered; in one case the solid solution
experiences a shallow quench in the metastable region, in another, it
is quenched deeply into the unstable region of the miscibility gap
\cite{Wagner_Kampmann_1991}. In the former case, the instability is
due to the formation of stable nuclei via localized composition
fluctuations, requiring to surmount an energy barrier that is
characterized by an incubation period. This corresponds to a
metastable phase and the transformation is by nucleation and
growth. In the latter case, the instability is caused by the formation
of non-localized, small-amplitude, spatially extended composition
fluctuations that grow spontaneously in amplitude as the time after
the quench proceeds \cite{Binder_1991}.  For a binary system, this
phenomenon is called spinodal decomposition, for which the quenched
state lies within the boundary between unstable and metastable states,
the so-called spinodal curve.

Langer and Schwartz \cite{Langer_Schwartz_1980} have developed a
detailed theory describing the nucleation, growth and coarsening of
droplets in metastable, near-critical fluids. Wendth and Haasen
\cite{Wendt_Haasen_1983} adapted this theory for solid solutions and
used it to interpret the precipitation of $\gamma '$-Ni$_3$Al SPPs in
Ni-14 at.$\%$Al. Later, Kampmann and Wagner \cite{Kampmann_Wagner_1984}
introduced a non-linearized Gibbs-Thomson relation into the
Langer-Schwartz equations in order to account for cases where the
surface tension of precipitates and the supersaturation are large.
They also included an empirical time-dependent factor on the
steady-state nucleation rate to include transient nucleation kinetics
that has been observed in many experiments. Kampmann and Wagner used
the model to interpret and explain observations on $\beta '$-Cu$_4$Ti
particles in Cu-1.9 at.$\%$Ti alloy.

In the present paper, we consider homogeneous nucleation, growth and
coarsening of SPPs under non-isothermal conditions. The
Langer-Schwartz equations for the precipitation kinetics are
formulated in a way that can be used to calculate nucleation and
growth of SPPs under time-varying temperature situations such as
quenching. A field-theoretic steady-state nucleation rate model is
used in our evaluation. Moreover, the basic models for growth and
coarsening are presented, followed by a computation of an isothermal
experiment on a copper-cobalt alloy phase transformation, and a
simulation of a body-centered cubic (bcc) to hexagonal close-packed
(hcp) phase quenching and subsequent annealing in the hcp phase of a
dilute zirconium alloy.

\section{Models}
\label{sec:model}

The models considered here describe the phenomena of homogeneous
nucleation, growth and coarsening of second phase precipitates in
non-isothermal conditions in an integrated fashion. The precipitates
are assumed to be spherical and at each instant are in local
equilibrium with their surrounding; therefore the concentration near
the SPP/matrix interface is determined by the Gibbs-Thomson
boundary condition. A key variable defining the state of the system is
the supersaturation; here defined as $x \equiv \ln{(C/C_\infty)}$,
where $C$ is the solute concentration in the matrix, which is
time-dependent, and $C_\infty$ the solubility limit or the equilibrium
solute concentration, which is strongly temperature dependent. The
Langer-Schwartz theory of precipitation basically comprises a
nucleation model and a growth model. In this section we outline these
models and describe our formulation of the theory for non-isothermal
applications.

\subsection{Nucleation}
\label{sec:nuke}
\subsubsection{Classical Theory}
\label{sec:cnt}
Consider that a spherical (or circular) particle with radius $R$ is
emerged from a metastable state. The particle free energy consists of
the surface and the bulk energy parts,
\begin{equation}
  \label{eq:spp_energy}
  F(R) = S_d\Big(\sigma R^{d-1}-\frac{\mu_e}{d} R^d\Big),
\end{equation}
\noindent
where $d$ is the spatial dimensionality, $S_3=4\pi$ and $S_2=2\pi$,
$\sigma$ the surface energy, and $\mu_e$ the difference in free
energy density (per unit volume) between the metastable and stable
phases. The function $F(R)$ goes through a maximum at a critical
radius $R=R_c=(d-1)\sigma/\mu_e$. The height of the nucleation energy
barrier is given by:
\begin{equation}
  \label{eq:spp_energy_max}
  F_c=F(R_c) = \frac{S_d}{d}\sigma R_c^{d-1}.
\end{equation}

The probability of formation of a particle of radius $R$ is $P\sim
\exp{[-\beta F(R)]}$, where $\beta=1/k_BT$ with $k_B$ Boltzmann's
constant and $T$ the temperature. Thus the critical nucleus, which
maximizes $F$, is the least probable one. Once the particle forms, it
will grow to reduce its free energy.

The basic ingredient of the considered model is the nucleation rate of
the secondary phase precipitates. In the standard formulation,
attributed to Ya. B. Zel'dovich \cite{Landau_Lifshitz_1981}, the
nucleation process is described by a Fokker-Planck equation:
\begin{equation}
  \label{eq:FPE}
  \frac{\partial n}{\partial t}=\frac{\partial}{\partial
R}\Big(B\frac{\partial n }{\partial R}\Big) + \beta \frac{\partial}{\partial
R}\Big(B n\frac{ \partial F}{\partial R}\Big),
\end{equation}
\noindent
where $n=n(R,t)$ is the number of SPP in the radius interval $[R,
R+dR]$ per unit volume at time $t$, $B=B(R)$ a nucleation size
diffusion coefficient and $F$ the SPP free energy. After a transient
time $t_0$, a constant nucleation rate of SPPs with radii larger than
$R_c$ prevails, $J=\mathrm{const}$.  More concretely, if $R>>R_c$,
$n(R,t)dR=Jdt$ is the number of growing newly emerging particles in a
time interval $dt$ per unit volume and is independent of $R$
\cite{Onuki_2002}.  The steady-state solution of Eq. (\ref{eq:FPE}),
$n=n_s(R)$ is obtained by integration, viz.,
\begin{equation}
  \label{eq:sss}
  n_s(R) = J\, e^{-\beta F(R)}\int_R^\infty dr
  B(r)^{-1}e^{\beta F(r)},
\end{equation}
\noindent
where the condition $n_s(R)\rightarrow 0$ as $R\rightarrow \infty$ was
imposed. Next, noting that the steady-state solution of the
Fokker-Planck equation is determined by the Boltzmann distribution,
i.e., $n_s=n_\ell\exp{[-\beta F(R)]}$, an expression for $J$ can be
obtained, when $R$ is near $R_c$, giving (Appendix \ref{sec:appenda})
\begin{equation}
  \label{eq:cnt2}
  J=(2\pi)^{-1/2}(d-1)^{1/2}D_n(S_d\beta\sigma)^{-1/2}
  R_c^{-(d+1)/2} n_l\exp{(-\beta F_c)},
\end{equation}
\noindent
where $D_n=S_d\beta\sigma R_c^{d-1}B_c$ is the critical nucleation
diffusivity with $B_c=B(R_c)$. Relating the critical radius to the
supersaturation $x$ and the capillary length $\ell$ according to:
$R_c=\ell/x$, where $\ell=2\beta\sigma v_a$ and $v_a$ the atomic volume of
the precipitate, Eq. (\ref{eq:cnt2}) is expressed in terms of the
supersaturation, i.e.,
\begin{equation}
  \label{eq:cnt3}
  J=\sqrt{\frac{d-1}{2\pi d}}D_n n_\ell\Big(\frac{x_0}{\ell}\Big)
  \Big(\frac{x}{x_0}\Big)^{\frac{d+1}{2}}\exp{\big[-\big(\frac{x_0}{x}\big)^{d-1}\big]},
\end{equation}
\noindent
where
\begin{equation}
  \label{eq:cap_length}
  x_0=\Big(\frac{S_d\beta\sigma}{d}\Big)^{\frac{1}{d-1}}\,\ell.
\end{equation}
\noindent
Relation (\ref{eq:cnt3}) is the classical expression for the
nucleation rate formulated in $d$ spatial dimensions as a function of
the supersaturation. The scaling in the critical region suggests that
the Boltzmann prefactor is $n_\ell \sim \ell^{-(d+1)}$. Whence, setting
$\mathcal{A}\equiv\ \sqrt{(d-1)/2\pi d}$, Eq. (\ref{eq:cnt3}) can be
written as
\begin{equation}
   J = \mathcal{A}\frac{D_nx_0}{\ell^{d+2}}\left( \frac{x}{x_o} \right)^{\frac{d+1}{2}}
\exp{\big[{- \left( \frac{x_o}{x}\right)^{d-1}}\big]}.
\label{eq:nr_bdz}
\end{equation}

\subsubsection{Field Theoretic Approach}
\label{sec:ftn}
The nucleation rate in the field-theoretic formulation
\cite{Langer_1969} begins from the relation that links the
steady-state nucleation rate with the imaginary part of the free
energy
\begin{equation}
  \label{eq:nr_gen}
J = \frac{\beta\kappa}{\pi}\,\Im\big[F(\mu_e)\big],
\end{equation}
\noindent
where $\kappa$ is a kinetic factor related to the diffusivity via
$\kappa=(d-1)D_n\ell/R_c^3$. Relation (\ref{eq:nr_gen}) has been
evaluated by various investigators and more carefully by G\"{u}nther, Nicole and
Wallace \cite{Gunther_Nicole_Wallace_1980}, which for $d=3$ takes the
form (Appendix \ref{sec:appendb})
\begin{equation}
J = \mathcal{B}\frac{D_nx_0^6}{\ell^5}\left( \frac{x}{x_o} \right)^{2/3}
\exp\big[{- \left( \frac{x_o}{x}\right)^2\big]},
\label{eq:nr_gnw}
\end{equation}
\noindent
where $\mathcal{B}=6^5/(288\pi\sqrt{3})$. Equation (\ref{eq:nr_gnw})
is strictly valid for low supersaturations, i.e., when $x/x_0 << 1$.
Langer and Schwartz \cite{Langer_Schwartz_1980} have heuristically
extended this expression for large values of $x$ according to
\begin{equation}
J = \mathcal{B}\frac{D_nx_0^6}{\ell^5}\left( \frac{x}{x_o} \right)^{2/3}
\left( 1+\frac{x}{x_o}\right)^{3.55}
\exp{\big[- \left( \frac{x_o}{x}\right)^2\big]}.
\label{eq:nr_ls}
\end{equation}

\begin{figure}[htbp]
  \centering \includegraphics[width=0.60\textwidth]{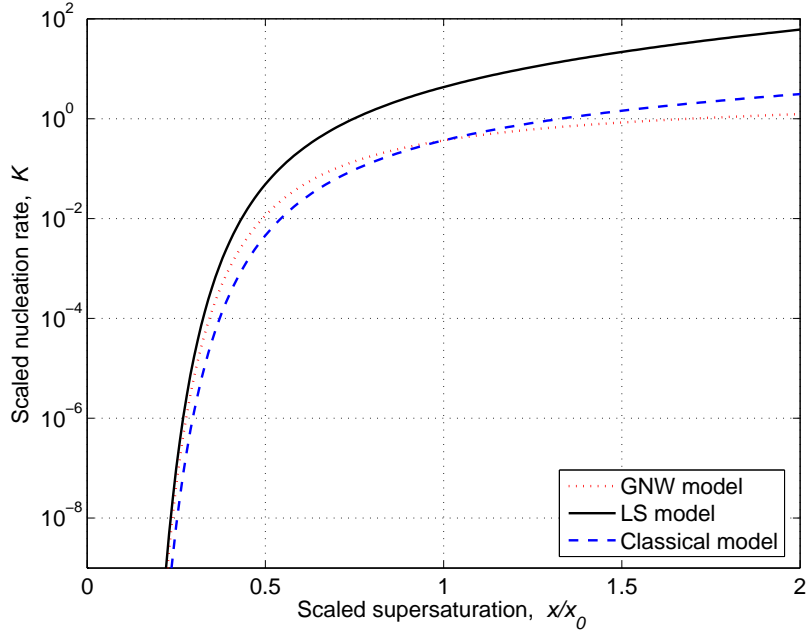}
    \caption{The scaled nucleation rate vs. scaled  supersaturation
      for $d=3$, comparing predictions of Eq. (\ref{eq:nr_bdz}),
      the classical  model,  Eq. (\ref{eq:nr_gnw}), the GNW model,  and
      Eq. (\ref{eq:nr_ls}), the LS  model. The scaling factors used are
      defined in the text}
    \label{fig:nucrate}
\end{figure}

The behavior of the aforementioned models for the nucleation rate, Eqs.
(\ref{eq:nr_bdz}), (\ref{eq:nr_gnw}) and (\ref{eq:nr_ls}) are
illustrated in Fig. \ref{fig:nucrate} for $d=3$, where the scaled
nucleation rate $K$ is plotted against the scaled supersaturation
$x/x_0$ (Eq. (\ref{eq:nr_bdz}) is scaled by
$\mathcal{A}D_nx_0/\ell^{d+2}$, while Eqs. (\ref{eq:nr_gnw}) and
(\ref{eq:nr_ls}) are scaled by $ \mathcal{B}D_nx_0^6/\ell^5$).
\subsection{Growth and Coarsening}
\subsubsection{Theory}
\label{sec:gct}
The details of the Langer and Schwartz theory have been clearly
described in their original work \cite{Langer_Schwartz_1980} and
reviewed in a number of publications (\cite{Wagner_Kampmann_1991},
\cite{Gunton_et_al_1983}, \cite{Aaronson_LeGoues_1992},
\cite{Ratke_Voorhees_2002}).  We only summarize its basic ingredients
in order to fix notation.  The basic equations of the Langer-Schwartz
theory are those of the Lifshitz and Slyozov
\cite{Lifshitz_Slyozov_1961}, supplemented by a source of droplets
(spherical precipitates) which describes the nucleation event. The
continuity equation for the droplet distribution function $n=n(R,t)$
is
\begin{equation}
  \label{eq:lasw-1}
 \frac{\partial n}{\partial t}+\frac{\partial}{\partial R}
 \Big [\frac{dR}{dt}n\Big]=j(R),
\end{equation}
\noindent
where $j(R)$ is the distributed nucleation rate of precipitates, $
J=\int_{R_c}^\infty j(R)dR$, and as before $J$ stands for the
nucleation rate per unit volume. Langer and Schwartz made a
simplifying assumption that only precipitates with $R>R_c$ are to be
counted as part of the second phase. The SPP number density $N$ 
is given by $ N=\int_{R_c}^\infty n(R)dR$ with their mean radius
defined as $ \overline{R}=\frac{1}{N}\int_{R_c}^\infty n(R)RdR$.

The growth rate of the spherical nucleus is controlled by the rate the
solute atoms supplied to the precipitate/matrix interface via
diffusion \cite{Lifshitz_Slyozov_1961}
\begin{equation}
\frac{dR}{dt} = \frac{D}{R} \left( \frac{C(t)-C_R}{C_p-C_R} \right).
\label{eq:Growth_rate}
\end{equation}
Here, $D$ is the diffusivity of the solute in the matrix, $C$ the
matrix average solute concentration, $C_p$ the solute concentration
within the precipitate, assumed to be uniform, and $C_R$ the local
solute concentration at the precipitate/matrix interface. The latter
concentration is related to the solubility limit $C_{\infty}$, a
temperature dependent quantity, and the curvature of the
precipitate/matrix interface through the Gibbs-Thomson relation $C_R =
C_{\infty} \exp{(\ell/R)}$. Also, the equation for conservation of matter must
be satisfied, viz.,
\begin{equation}
\frac{C-C_0}{C-C_p} = \frac{4\pi}{3} \overline{R}^3N(t).
\label{eq:conserve}
\end{equation}
\noindent

For non-isothermal conditions, where temperature is varying with time,
all the temperature dependent parameters appearing in the foregoing
equations must be considered and treated properly. The resulting
Langer-Schwartz differential equations for $N$ and $\overline{R}$,
applicable to both isothermal and non-isothermal conditions, can be
expressed by
\begin{equation}
\frac{dN}{dt} = \frac{g_1-g_1f_2+g_3f_1}{1-f_2-g_2g_3f_3}; \quad 
\frac{d\overline{R}}{dt} = \frac{f_1 -f_1g_2+f_3g_1}{1-g_2-f_2f_3g_3},
\label{eq:dNdt_final}
\end{equation}
\noindent
where the functions $g_1, g_2, g_3$ and $f_1, f_2, f_3$ are
\begin{eqnarray}
g_1 & = & J + \frac{ b \ell N }{ x (\overline{R}-R_c) }
\left( \frac{1}{T} - \frac{1}{x} 
\frac{d \ln{C_{\infty}}}{d T} \right)
\frac{dT}{dt},
\label{eq:g_1} \\
g_2 & =  & \frac{ -N \overline{R} }{ 3(\overline{R}-R_c) }\mathit{\Gamma},
 \qquad g_3  =  \frac{ -N^2 }{ (\overline{R}-R_c) }  \mathit{\Gamma},
\label{eq:g_2}
\end{eqnarray}
\begin{eqnarray}
f_1 & = & \frac{D}{\overline{R}}
\left( \frac{ e^x C_\infty-C_{\overline{R}}}{C_p-C_{\overline{R}}} \right) +
\frac{J}{N} \left( R_c + \delta R_c - \overline{R} \right) + \frac{b\ell}{x} \left( \frac{1}{x} 
\frac{d\ln{C_{\infty}}}{d T} - \frac{1}{T} \right) \frac{dT}{dt},
\label{eq:f_1} \\
f_2 & = & N  \mathit{\Gamma},
 \quad f_3  =  \frac{\overline{R}}{3}  \mathit{\Gamma},  \quad
\mathit{\Gamma} =  \frac{4\pi b \ell \overline{R}^2(C_p-C_0)}
{C_\infty e^x x^2 \big(1-\frac{4\pi\overline{R}^3}{3}N\big)^2}.
\label{eq:f_2}
\end{eqnarray}
In the above expressions, $\delta R_c$ is the width of the size
distribution and $b=0.317014$ for $\overline{R} \le 3R_c/2$ and $b=0$
otherwise \cite{Langer_Schwartz_1980}. In computations, we assume
$\delta R_c= a \ell$ with $a$, a positive constant $a<1$, taken as a free
parameter. It is noted that the particular value of $b$ set by Langer
and Schwartz reproduces the familiar Lifshitz-Slyozov coarsening law
``$t^{1/3}$'' as $t \rightarrow \infty$ \cite{Langer_Schwartz_1980,
  Lifshitz_Slyozov_1961}. The initial conditions for the equations in
(\ref{eq:dNdt_final}) are
\begin{eqnarray}
N(t=t_o) & = & N_o, \label{eq:N_o} \\
\overline{R}(t=t_o) & = & R_c + \delta R_c = \frac{\ell}{x} + a\ell, \label{eq:R_o} 
\end{eqnarray}
\noindent
where the starting time $t_o$ is determined from $N_o = \int_{0}^{t_o}
J(\tau) d\tau$; and $N_o$ defines the lowest particle density of practical
interest, here treated as a model parameter.
\subsubsection{Numerical Method}
The nonlinear ordinary differential equations (\ref{eq:dNdt_final})
need to be evaluated numerically. In evaluations of the
right-hand-side expressions of equations in (\ref{eq:dNdt_final}), we
make use of correlations for the solubility limit $C_{\infty}$ and the
diffusivity $D$, which are temperature-dependent material
properties. We have used the MATLAB programming environment to solve
these differential equations; more specifically the MATLAB solver
\mbox{ODE15S} \cite{Matlab}. This is a variable order solver, which is
intended to solve stiff systems of ordinary differential equations.
The solver is invoked at each time step of the time-temperature
history, and the evolution of $N$ and $\overline{R}$ during the time
step is computed with ODE15S for each mesh point, based on the local
temperature at beginning and end of the time step.  In the solution
procedure, the temperature is assumed to vary linearly with time
during the time step.

\section{Application}
 \label{sec:exp}
\subsection{Isothermal experiment}
\label{sec:iso}

The methods outlined in the foregoing section have been verified
\cite{Jernkvist_2005} against a number of measurements made in
isothermal conditions for binary precipitates such $\gamma '$-Ni$_3$Al
in Ni-14 at.$\%$Al and Co in Cu-Co alloys
\cite{Wendt_Haasen_1983,LeGoues_Aaronson_1984, Wendt_Haasen_1985,
Haasen_Piller_1987}. Here, we only report the results of our
computations for cobalt precipitates in Cu-2.7 at.\%Co alloy, for
which isothermal annealing tests (at 823 K) were carried out by Wendth
and Haasen \cite{Wendt_Haasen_1985}. The considered experiment
involved measurements of the SPP number density as a function of
annealing time and also the time variation of precipitate mean size
was determined. These properties were measured by transmission
electron microscopy (TEM) and atom probe field ion microscopy.

In order to simulate this test, we used the solubility limit for
cobalt in copper proposed by Servi and Turnbull \cite{Servi_Turnbull}
$C_{\infty} = 712.85 \times 10^{-2875/T}$, where $T$ is the absolute
temperature and $C_{\infty}$ is in atomic percent. For the diffusivity
of cobalt in copper, we used the correlation by D\"{o}hl \textit{et al.}
\cite{Dohl_etal}, $D = 4.3 \times 10^{-5} e^{-25738/T}$, where $D$ is
in m$^2$s$^{-1}$.  The matrix/SPP interface energy $\sigma$ was in our
analyses set to 0.22 Jm$^{-2}$, following the evaluation of data by
Stowell \cite{Stowell}.  The second phase particles in these alloys
are of pure cobalt, which means that $C_p$ is 100 at\%. Both the
particles and the matrix have a face-centered cubic (fcc) crystal
structure. Moreover, the molar volume $v_m$ of cobalt is $6.7 \times
10^{-6}$ m$^{3}$mole$^{-1}$. In analyses, we have set the model
parameter $N_o$ in Eq. (\ref{eq:N_o}) to $1 \times 10^{10}$ m$^{-3}$
and $a$ in Eq. (\ref{eq:R_o}) to unity.
\begin{figure}[htbp]
  \begin{center}
    \includegraphics[width=0.60\textwidth]{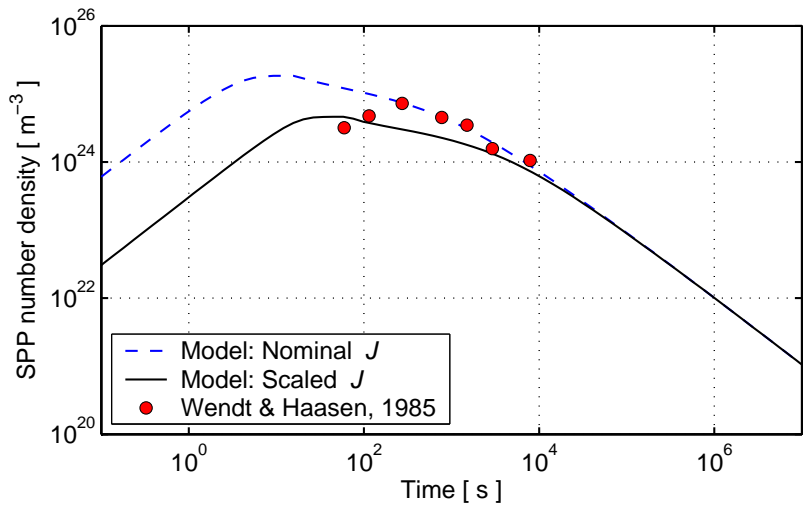}
    \caption{Calculated number density of Co second phase particles in
             Cu-2.7at\%Co at 823 K, in comparison with data from 
             \cite{Wendt_Haasen_1985}. 
             The calculated nucleation rate, given by Eq. (\ref{eq:nr_ls}), is
             scaled by a factor 0.05 in order to fit the data.}
    \label{fig:CuCo_T823_Nnom}
  \end{center}
\end{figure}

The calculated evolution of SPP number densities are compared with
data in Fig. \ref{fig:CuCo_T823_Nnom}. The results are shown for two
cases: For the nominal case, the nucleation rate $J$, given by
Eq. (\ref{eq:nr_ls}), has been used.  For the scaled case, the
coefficient $\mathcal{B}$ in Eq. (\ref{eq:nr_ls}) has been scaled by a
factor 0.05, in order to obtain a best fit to the data. For this test
at 823 K, shown in Fig. \ref{fig:CuCo_T823_Nnom}, the measured SPP
density reaches a peak at about 300 seconds, after which coarsening
starts to dominate the picture. At this stage, large particles will
grow at the expense of smaller ones, as a consequence of the
Gibbs-Thomson relation, see section \ref{sec:gct}. The smallest
particles will ultimately dissolve, which makes the number density
decrease with a rate proportional to $\approx 1/t$.  Our model
captures this behavior fairly well, but the calculated peak in the SPP
density is reached too early, in comparison with data. For Cu-1.0
at\%Co alloy, LeGoues and Aaronson \cite{LeGoues_Aaronson_1984}
reported a delay time of about 130 s in their study, whereas Haasen
and Piller \cite{Haasen_Piller_1987} reported that the delay time was
no more than 30 s in the study made by Al-Kasab.  Since we have
utilized a steady-state theory and assumed a homogeneous nucleation
rate, we could have overlooked the effect of the delay time, $t_d$,
and hence get a discrepancy between the observed temporal data and
calculated values; cf.  Appendix \ref{sec:appendc} for a rough
estimation of this effect.

\subsection{Non-isothermal simulation}
\label{sec:noniso}

\begin{figure}[htbp]
  \begin{center}
    \includegraphics[width=0.60\textwidth]{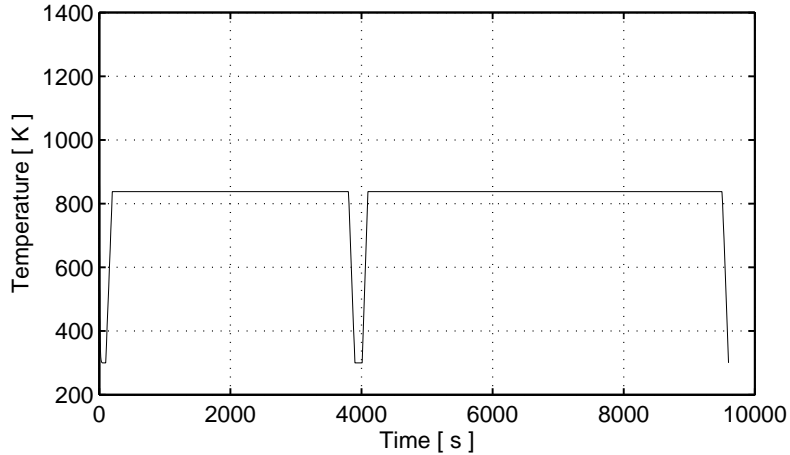}
    \caption{Temperature history for the annealing steps; the cooling and heating
rates for the steps are $\pm 5.4$ Ks$^{-1}$.}
    \label{fig:t-hist}
  \end{center}
\end{figure}

In this subsection, we present the results of a simulation of a heat
treatment made on a cylindrical specimen.  The cylinder is made of
Zircaloy-2 (Zr-1.4Sn-0.12Fe-0.09Cr by wt.\%) and has a diameter of 25
mm. The heat treatment consists of quenching the sample
from a high temperature (1323 K) body-centered cubic $\beta$ phase in
water, which is kept at room temperature. After the quenching, the
cylinder is subjected to two subsequent annealing steps in Ar gas at
838 K (hexagonal-closed packed $\alpha$ phase) for durations of 1.0 h
and 1.5 h, respectively (Fig. \ref{fig:t-hist}).
The prominent second phase precipitate in Zircaloys is Zr(Fe,Cr)$_2$
(Laves phase with hexagonal structure) formed at around 1200 K
\cite{Massih_et_al_2003}. Here, we calculate the nucleation and growth
of this kind of precipitates under the aforementioned heat treatment
(quenching plus annealing) using the models described in the foregoing
section, which are implemented in a computer program
\cite{Jernkvist_Massih_2004}. In particular, we have selected the
nucleation rate equation (\ref{eq:nr_ls}) in our evaluation. Since the
diffusivity of Fe and Cr are similar and solubility data (temperature
vs. concentration) are available for (Fe+Cr) in Zircaloy
\cite{Massih_et_al_2003}, we have treated the precipitate as a binary
compound Zr and (Fe+Cr) with initial solute (Fe+Cr) concentration of
$C_0=2010$ wppm (weight parts per million). The SPP molar volume is
$v_m=9 \times 10^{-6}$ m$^{3}$mole$^{-1}$ and its composition $C_p =
5.4\times 10^5$ wppm.  We have assumed that the diffusivity during
nucleation is identical to that during growth. The input model
parameters in our evaluations are as follows. We set the effective
diffusivity for (Fe+Cr) in Zircaloy as $D(T) = 1.473 \times 10^{-6}
e^{-15930/T}$ , where $T$ is the absolute temperature and $D$ is in
m$^2$s$^{-1}$, and the surface tension of the precipitate
$\sigma=0.25$ Jm$^{-2}$. The lower cut-off limit for the particle
density, $N_o$ in Eq. (\ref{eq:N_o}), is set to $1 \times 10^{10}$
m$^{-3}$, and the model parameter $a$ in Eq.  (\ref{eq:R_o}) to 0.25.
The motivations for selecting these values are discussed in
\cite{Jernkvist_2005}.

\begin{figure}[htbp]
  \begin{center}
    \begin{tabular}{ll}
      \includegraphics[width=0.50\textwidth]{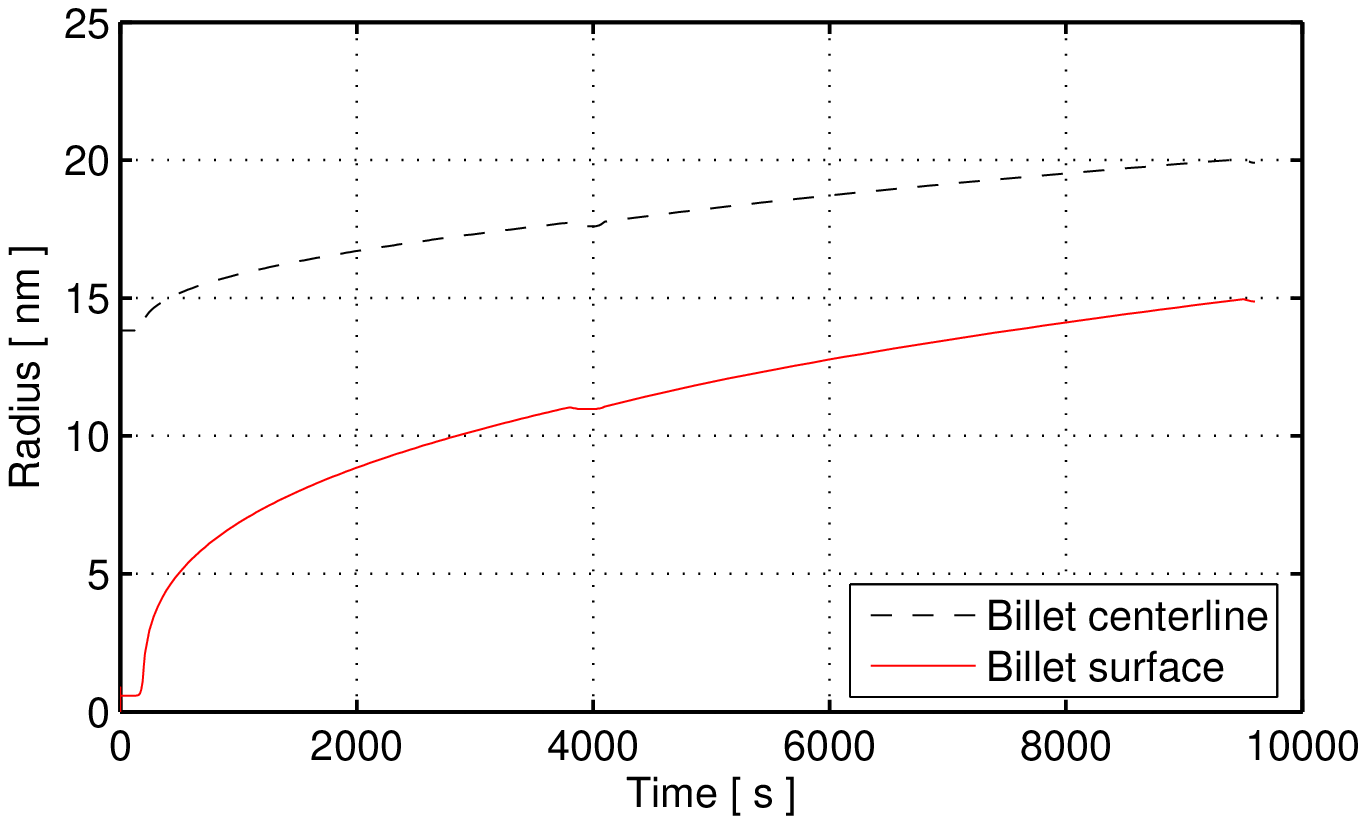} &
      \includegraphics[width=0.50\textwidth]{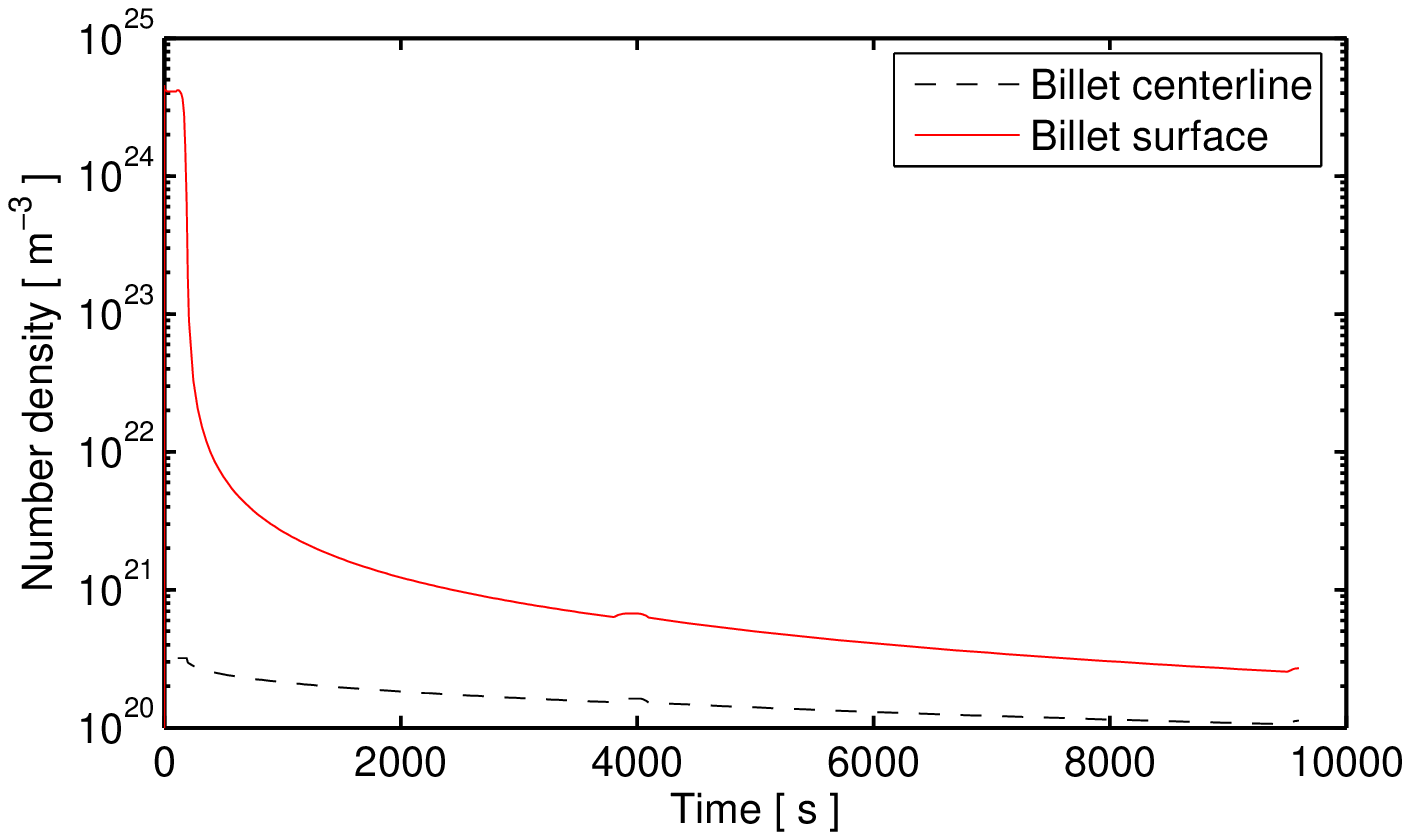} \\

       \includegraphics[width=0.50\textwidth]{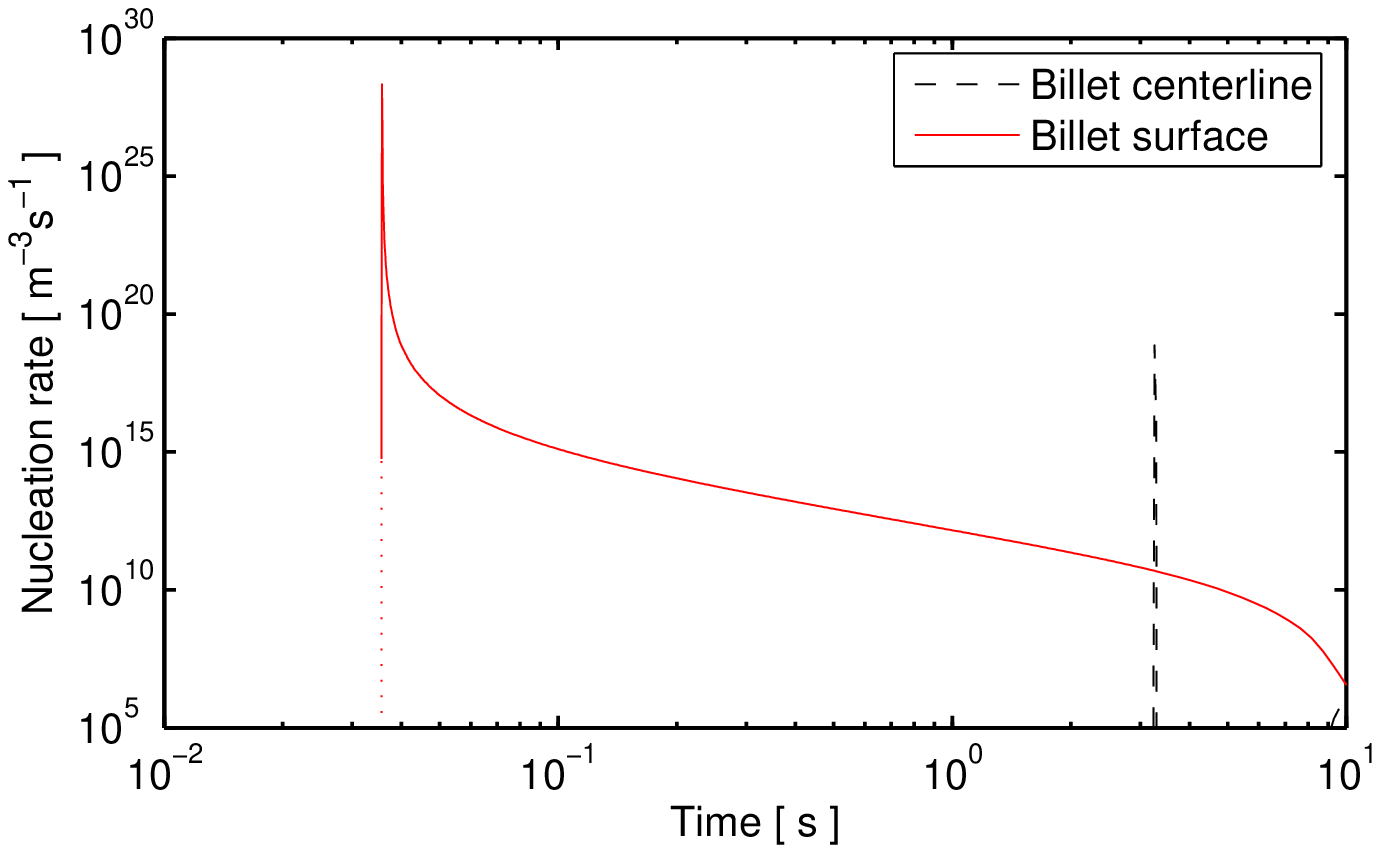} &
      \includegraphics[width=0.50\textwidth]{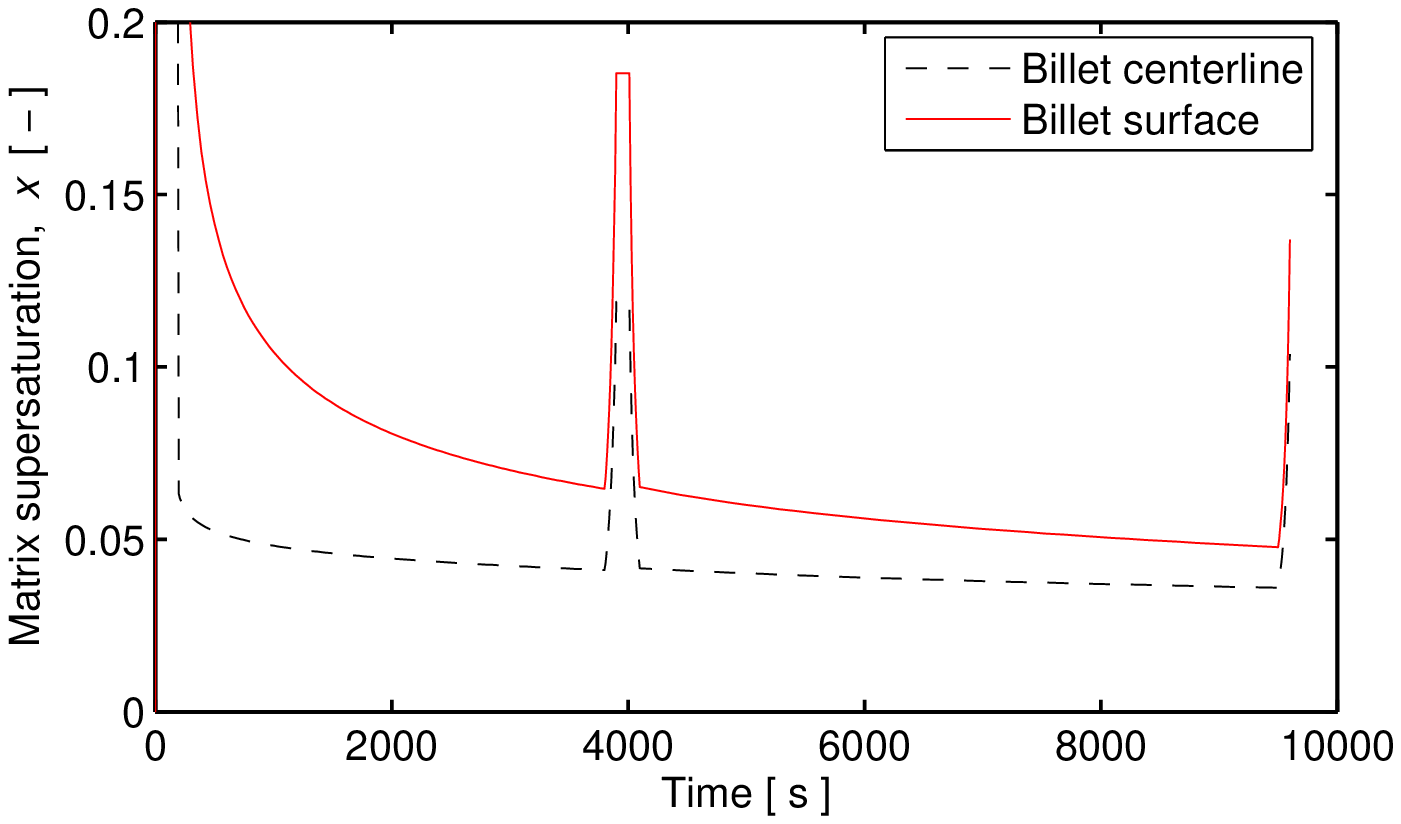} \\

      \includegraphics[width=0.50\textwidth]{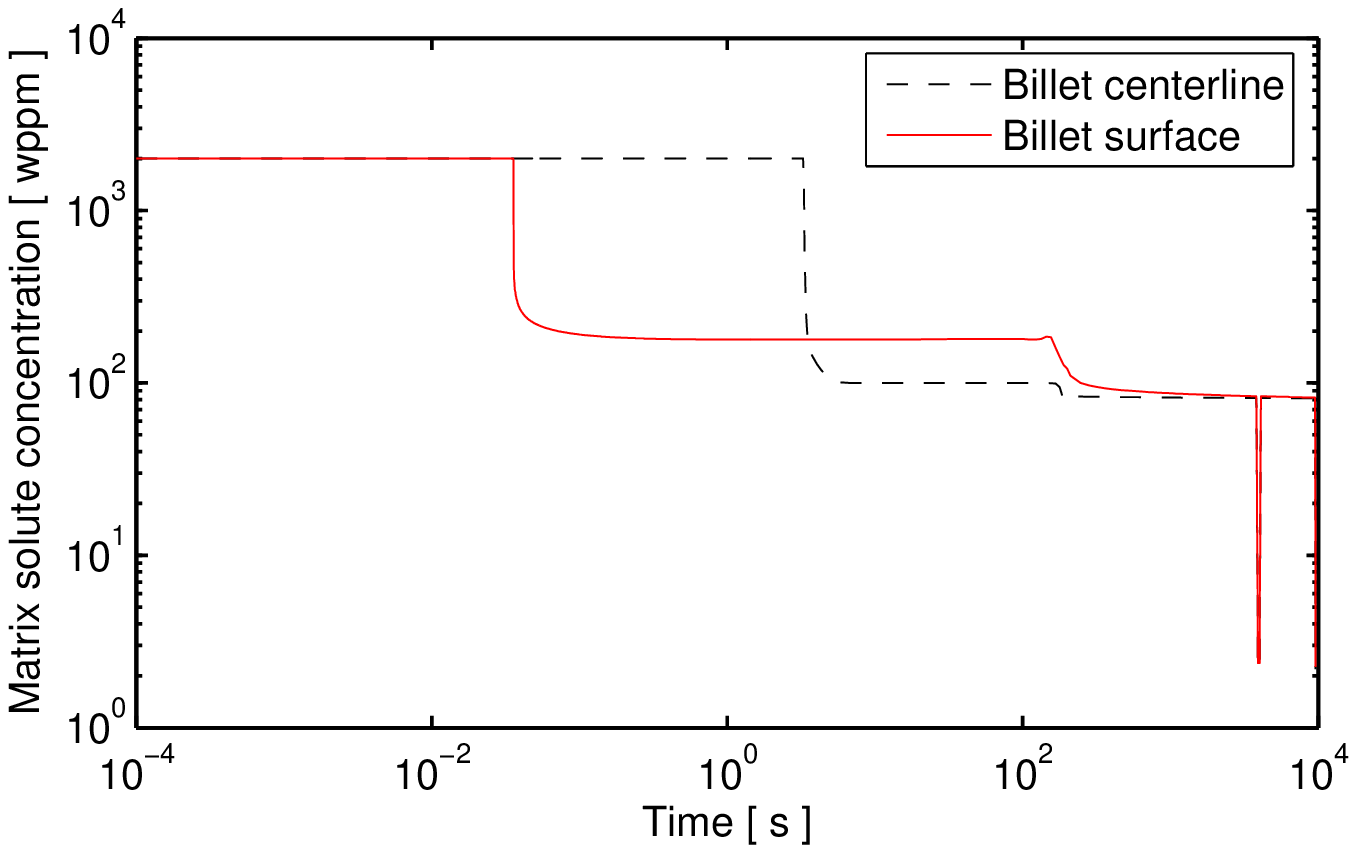} &
      \includegraphics[width=0.50\textwidth]{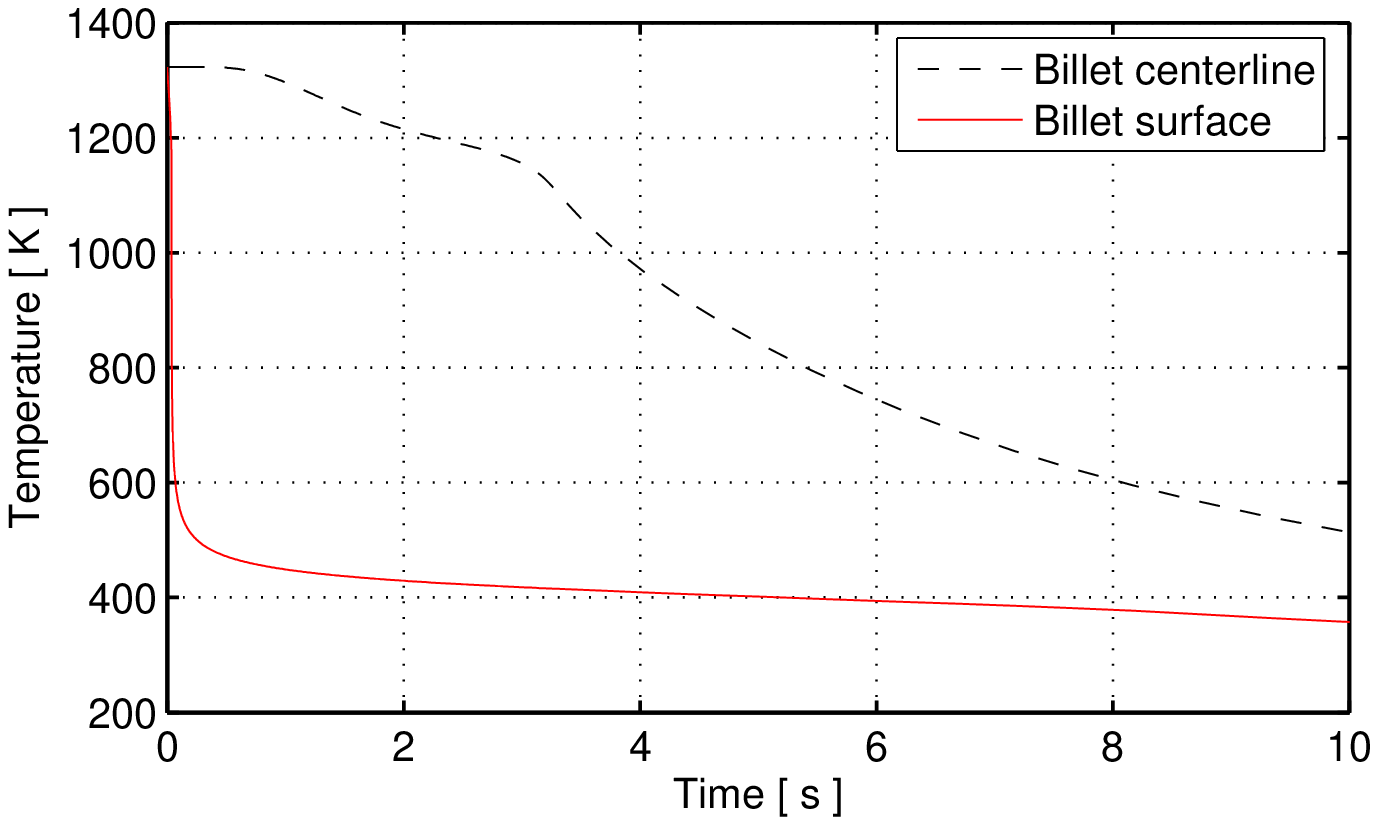} \\
      
      \end{tabular}
    \caption{Evolution of Zr(Fe,Cr)$_2$ SPPs in a
      Zircaloy-2 cylinder during the heat treatment. Diagrams, left to
      right, top to bottom, show calculated (i) mean precipitate radius,
      (ii) precipitate number density, (iii) nucleation rate, (iv) matrix
      supersaturation, (v) matrix solute concentration (Fe+Cr), and
      (vi) temperature in the first ten seconds.}
    \label{fig:spp-evol}
  \end{center}
\end{figure}

The results of our non-isothermal evaluation are presented in a number
of diagrams in Fig. \ref{fig:spp-evol}, which shows (i) the mean
precipitate radius, (ii) the precipitate number density, (iii)
nucleation rate, (iv) the matrix supersaturation, (v) the matrix
solute concentration (Fe+Cr), and (vi) temperature evolution in the
first ten seconds of the heat treatment. The computation output
corresponds to the center of the cylinder and 0.1 mm from the outer
surface. The cooling rates at these locations, during
quenching, can be noted from the temperature vs. time diagram in
Fig. \ref{fig:spp-evol}.

\subsection{Discussion}
\label{sec:disc}
From the results displayed in Fig.  \ref{fig:spp-evol}, we note that
there is a significant difference in calculated precipitate radius at
the different positions in the cylinder, directly upon $\beta$
quenching, but this difference gradually diminishes under subsequent
annealing.  The initial difference in the radius of the precipitates
located in the cylinder central region and those close to the outer
surface follows from the difference in the cooling rate, as is seen in
the temperature diagram in Fig.  \ref{fig:spp-evol}.  According to our
computations, the slow cooling at the cylinder central part results in
appreciable precipitate growth already under the quenching phase. This
behavior is not predicted at the billet outer surface.

The calculated time variation of the precipitate number density shown
in Fig.  \ref{fig:spp-evol} displays a remarkable difference between
the central and periphery of the cylinder directly on quenching. The
calculated number density at the outer surface is initially about
several orders of magnitude larger than that at the central part, but
this large difference gradually disappears under subsequent annealing
as a result of coarsening. Figure \ref{fig:spp-evol} also shows the
calculated nucleation rate as a function of time, evaluated at the
considered locations in the cylinder.  As can be seen, the nucleation
takes place under a very short time span, as the temperature drops
below 1118 K and the matrix becomes supersaturated ($C>C_{\infty}$).
The nucleation rate pulse width, as can be seen in the diagram, is a
couple of ms at the billet outer surface, whereas the corresponding
time span is much larger in the central part, i.e., around 50 ms.  The
calculated matrix supersaturation, $x = \ln{(C/C_\infty)}$, is also
shown as a function of time in the figure.  The spikes seen in the
diagram for $x$ correspond to the temperature drops during the
annealing cycles (Fig. \ref{fig:t-hist}), which reduce
$C_\infty$. Note also the sharp dive in the matrix solute
concentration due the temperature dip. The results of the computations
presented here, i.e., the precipitate radius and number density, are
in qualitative agreement with the observations made on a similar kind
of heat treatment of this material \cite{Massih_etal_CS}.

As alluded in section \ref{sec:iso}, we have supposed a steady-state
nucleation model. Such a treatment does not provide information
regarding instantaneous SPP size distribution nor on the nucleation
rate prior to reaching steady state. Some workers postulate a
nucleation relation in the form $J(t)=J(\infty)\exp{(-t_d/t)}$ to fit
metallurgical data \cite{Wagner_Kampmann_1991}. However, this type of
approach is simplistic and a more rigorous treatment to solve the
time-dependent Fokker-Plank equation (\ref{eq:FPE})
\cite{Kashchiev_1999} or extend the field theoretical approach
(section \ref{sec:iso}) to the realm of phase transition dynamics can
be more expedient. Besides, nucleation does not commonly occur
homogeneously in a matrix by means of solely thermal and concentration
fluctuations. It is induced most often by heterogeneities and defects
in solids, e.g., grain and interphase boundaries, dislocations,
stacking faults, free surfaces and vacancies or their clusters. These
micro-domains facilitate or enhance the nucleation rate for formation
of a new phase during quenching \cite{Porter_Easterling_1981}.

\section{Conclusion}
 \label{sec:concl}
 The equations for nucleation, growth and coarsening of second phase
precipitates are extended to account for the
non-isothermal situations. The results of our simulation of the heat
treatment of a Zircaloy-2 specimen clearly illustrates the influence
of cooling rate during quenching on the properties of SPPs. A lower
cooling rate results in larger precipitates with a smaller number
density. This calculation is in qualitative agreement with the
observations made on a similar kind of heat treatment of this material
\cite{Massih_etal_CS}, which also indicate that SPP size and density
are quenching rate dependent and they impact macroscopic properties of
the alloy.


\appendix
\section{Nucleation rate close to the critical droplet radius}
\label{sec:appenda}

When the droplet (precipitate) radius $R$ is close to its critical
value $R_c$, the free energy, $F(R)$ in Eq. (\ref{eq:spp_energy}) can
be approximated by, $F(R)=F_c-0.5(d-1)S_d\sigma R_c^{d-3}(\delta
R)^2$, with $\delta R=R-R_c$ \cite{Onuki_2002}. The integrand in Eq.
(\ref{eq:sss}) is sharply peaked at $R_c$, thus the steepest descent
method is used to evaluate the integral, resulting:
\begin{equation}
  \label{eq:cnt0}
  J=(2\pi)^{-1/2}B_c\Big(\beta\vert F^{\prime\prime}(R_c)\vert\Big)^{1/2} n_s(R_c),
\end{equation}
\noindent
Substituting $n_s(R_c)=n_l\exp{[-\beta F(R_c)]}$ and
simplifying
\begin{equation}
  \label{eq:cnt1}
  J = (2\pi)^{-1/2}\Omega_c\varepsilon R_c n_l\exp{(-\beta F_c)}.
\end{equation}
Here, $\Omega_c=B_c\beta \vert F^{\prime\prime}(R_c)\vert$ is the nucleation
frequency, $B_c=B(R_c)$ the critical kinetic coefficient for
nucleation and $\varepsilon=\big(\beta \vert
F^{\prime\prime}(R_c)\vert R_c^2\big)^{-1/2}$ with
$F^{\prime\prime}(R_c)=\big(\partial^2 F/\partial R^2\big)_{R=R_c}$.
Utilizing Eq. (\ref{eq:spp_energy}), we write, $\Omega_c =
S_dB_c\beta\sigma (d-1)R_c^{d-3}$ and $\varepsilon^{-2} =
S_d\beta\sigma(d-1)R_c^{d-1}$.  Placing these last two relations into
Eq. (\ref{eq:cnt1}), we obtain Eq. (\ref{eq:cnt2}) of the main text.

\section{Statistical Description of Nucleation Rate}
\label{sec:appendb}
\begin{figure}[htbp]
  \begin{center}
    \includegraphics[width=0.60\textwidth]{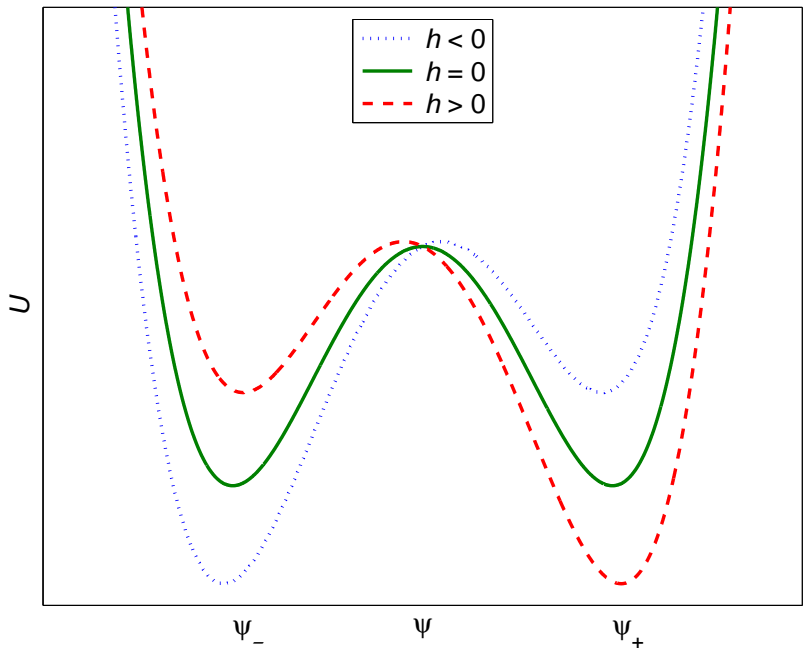}
    \caption{The potential energy $U$ with stable and metastable minima.}
    \label{fig:towell}
  \end{center}
\end{figure}

The starting point for the statistical mechanics approach to
nucleation theory is the Ginsburg-Landau free energy functional for a
scalar field variable $\psi(\vec{r}\,)$:
\begin{equation}
\label{eq:lge}
  \widetilde{F}(\psi)=\int d\vec{r}\, \Big[\frac{K}{2}(\nabla \psi)^2+U(\psi)\Big],
\end{equation}
\noindent
where $\psi=\psi(\vec{r}\,)$ is the order parameter, e.g., for a binary
alloy it is the local concentration, $K$ is a phenomenological constant
designating the amplitude of the spatial gradient of $\psi$ and $U$ is
the potential energy, which for a ``$\psi^4$'' field theory is
$U(\psi)=-\frac{r_0}{2}\psi^2+\frac{u_0}{4}\psi^4-h\psi$. Here the
coefficients $r_0$ and $u_0$ are, in general, functions of pressure and
temperature and $h$ is an acting external field, which for the case of
binary alloys is equivalent to the chemical potential. The potential
$U$ has two minima, $\psi_{\pm}(h)$. For $h>0$, $\psi_{+}$ expresses
the stable phase and $\psi_{-}$ the metastable phase, whereas for
$h<0$ the roles of $\psi_{+}$ and $\psi_{-}$ are reversed
(Fig. \ref{fig:towell}).

To proceed, consider the grand-canonical partition function for the
system, defined as:
\begin{equation}
\label{eq:zus}
  Z(h)=\int \delta\psi e^{-\beta \widetilde{F}(\psi)}.
\end{equation}
The extrema of the integrand in $Z$ are the solutions of the
equilibrium equation, i.e.,
\begin{equation}
\label{eq:ele}
  K\nabla^2 \psi+r_0\psi-u_0\psi^3+h=0.
\end{equation}
\noindent
One solution of interest, for small $h$, is the saddle point
solution, $\bar{\psi}$. It describes a radially symmetric particle of
the nucleating phase embedded in a spatially uniform  metastable
background. Its profile is approximated by
\begin{equation}
\label{eq:sps}
\bar{\psi} \approx \frac{1}{2}(\psi_+ + \psi_-) + \frac{1}{2}(\psi_+ -\psi_-)
\tanh{\Big[\frac{r-R}{\sqrt 2 \xi}\Big]},
\end{equation}
\noindent
where $\psi(r=0)=\psi_-$ and $\psi(r \to \infty)=\psi_+$ are the
stable and metastable values of the order parameter, corresponding to
the two minima of $U(\psi)$, describing a stable particle in a
metastable surrounding, $r$ is the radial position, $R$ the particle
radius and $\xi=(K/2r_0)^{1/2}$ is the interface thickness called the
correlation length. In a $\psi^4$ field theory
the correlation length is related to the capillary length $\ell$,
cf. section \ref{sec:nuke}, according to $\xi=6\ell$ \cite{Onuki_2002}.

 The free energy $\widetilde{F}(\psi)$ may be expanded around the
saddle point $\bar{\psi}$. Using Dirac's concise notation, we write
\begin{equation}
\label{eq:lgex}
  \widetilde{F}(\psi)=\bar{F}+\frac{1}{2}\langle u|\mathcal{M}|u\rangle+\mathcal{O}(u^3),
\end{equation}
\noindent
where $\bar{F}=F(\bar{\psi})$, $u=\psi-\bar{\psi}$, and $
\mathcal{M}=-K\nabla^2-U^{\prime\prime}(\bar{\psi})$.
We note that since $\bar{\psi}$ is a saddle point there is one
negative eigenvalue which reflects the instability of the critical
particle. Moreover, there are $d$ zero eigenvalues associated with the
translational modes of the particle in $d$-dimension. More explicitly,
as Langer \cite{Langer_1967} showed, the eigenvalues in the second
order in $h$ can be expressed by
\begin{equation}
\label{eq:evs}
  \lambda_{0l}=\frac{(l-1)(l+d-1)}{R_c^2}\Big[1+\mathcal{O}\big(\frac{l^2\xi^2}{R_c^2}\big)\Big],
\qquad l=0,1,2,\dots
\end{equation}
\noindent
where $R_c=(d-1)\sqrt{2K}r_0/(3h\sqrt{u_0})$ is the critical radius of the
nucleus. Here, $\lambda_{00}$ is the only negative eigenvalue, whereas
$\lambda_{01}$ is the $d$-fold degenerate zero eigenvalue. The energy
of the lowest mode $l=0$ is negative, implying that in the context of
a steepest descent calculation the critical particle generates the
imaginary part of the free energy in the metastable state.

As it has been pointed out in \cite{Gunther_Nicole_Wallace_1980}, in
the case of interest ($h \to 0$, $R_c \to \infty$), the eigenfunctions
of $\mathcal{M}$ form a band of soft modes which describe the surface
excitations of the spherical critical particle. In fact, the
eigenfunction corresponding to $l=1$ is the Goldstone mode associating
with the spontaneous breaking of translational invariance by the
center of the droplet. Apart from the unstable and translational modes
the remaining eigenvalues of $\mathcal{M}$, $\lambda_{nl}$ are positive
describing stable distortion of the critical spherical particle
\cite{Langer_1967}.

To calculate the nucleation rate, we first express the partition function
(\ref{eq:zus}) in the form of the linked-cluster expansion, viz.,
\begin{equation}
\label{eq:z-lce}
  Z=Z_0\exp\Big(\frac{Z_1}{Z_0}\Big),
\end{equation}
\noindent
where $Z_0$ represents the contribution from the metastable
minimum and $Z_1$ that from the path  $\bar{\psi}$. The contribution
from the metastable minimum $Z_0=\exp[-\beta F(\psi)]$, with
$\psi(r)=\hat{\psi}_+  + \eta(r)$ is written in the form
\begin{equation}
\label{eq:z-0}
  Z_0=\exp[-\beta F(\hat{\psi}_+)]\prod_j\Big(\frac{\pi}{\beta R^d \lambda_j^{(0)}}\Big)^{1/2},
\end{equation}
\noindent
where $ \lambda_j^{(0)}$ are the eigenvalues of the equation:
$\mathcal{M}_0\eta_j^{(0)}=\lambda_j^{(0)}\eta_j^{(0)}$ with
$\mathcal{M}_0=-K\nabla^2-r_0/2+1.5u_0\hat{\psi}_+^2$. Here
$\hat{\psi}_+=2(r_0/3u_0)^{1/2}\cos\theta$ where $3\theta
  =\arccos[1.5h(3u_0/r_0^3)^{1/2}]$ is the the stable solution of Eq.
  (\ref{eq:ele}) for $\nabla^2\psi=0$ and $h>0$. The saddle point
  contribution to the partition function can be written as
\begin{equation}
\label{eq:z-1}
  Z_1=\exp[-\beta F(\bar{\psi})] \mathcal{J}\mathcal{V}
\prod_{n,l}^{\prime}\Big(\frac{\pi}{\beta R^d \lambda_{nl}}\Big)^{(2l+1)/2},
\end{equation}
\noindent
where the prime denotes that the zero modes are removed and
$\mathcal{J}$ is the Jacobian that arises in integration and
$\mathcal{V}$ denotes the contribution of zero modes and is
proportional to the system volume V. The free energy of the system is
$\mathcal{F}=\lim_{V\to \infty} \beta^{-1}V^{-1}\ln Z$. Writing
Eq. (\ref{eq:z-lce}) in the form $\ln Z = \ln{Z_0 + Z_1/Z_0}$ and
using Eqs. (\ref{eq:z-0}) and (\ref{eq:z-1}), the droplet free energy
is expressed by
\begin{equation}
\label{eq:gnw-a5}
   \mathcal{F}=\exp[-\beta F(\bar{\psi})]
\mathcal{J}\mathcal{V}r_0^{d/2}\exp{\Big[\sum_{l \ne 1}^L\ln{\Big(\frac{r_0}{\lambda_{0l}}\Big)^{\nu_l/2}}\Big]},
 \end{equation}
\noindent
where $\nu_l$ is the degeneracy of the $d$-dimensional spherical
harmonics, e.g., for $d=2$, $\nu_l=2$ while for $d=3$,
$\nu_l=2l+1$. The upper limit of the sum $L$ is given by $L \sim
R_c^2r_0$, for which the approximation (\ref{eq:evs}) becomes
invalid. G\"{u}nther, Nicole and Wallace
\cite{Gunther_Nicole_Wallace_1980} have calculated the factors in
Eq. (\ref{eq:gnw-a5}). Placing $\rho \equiv
r_0^{3/2}/\big(|h|\sqrt{u_0}\big)$, these factors are
\begin{eqnarray}
F(\bar{\psi}) & = & c\frac{r_0^{(4-d)/2}}{u_0}\rho^{d-1},
\label{eq:gnw-a8} \\
\mathcal{J} & =  &  \rho^{(d-1)d/2}r_0^{(4-d)d/4}u_0^{-d/2},
\label{eq:gnw-a9}
\end{eqnarray}
\noindent
and the spectrum of excitations are calculated to be
\begin{equation}
\exp{\Big[\sum_{l \ne
1}^L\ln{\Big(\frac{r_0}{\lambda_{0l}}\Big)^{\nu_l/2}}\Big]} \propto
\left\{\begin{array}{ll} i\rho^{-d}\exp{\big(k_0\rho^{d-1}\big)} & d \ne 3
 \\ i\rho^{-2/3}\exp{\big(k_0\rho^{2}\big)} & d=3 \end{array}\right.
\label{eq:gnw-a12}
\end{equation}
\noindent
where $c$ and $k_0$ are dimensionless constants and $i=\sqrt{-1}$.
The nucleation rate \`{a} la Langer \cite{Langer_1969} is
$J=\beta\kappa\Im(\mathcal{F})/\pi$ with $\kappa$ being a kinetic
factor; Eqs. (\ref{eq:gnw-a8}), (\ref{eq:gnw-a9}) and
(\ref{eq:gnw-a12}) can be assembled according to
Eq. (\ref{eq:gnw-a5}) to yield the GNW relation
\cite{Gunther_Nicole_Wallace_1980} in the regime of interest ($h \to
0$, $R_c \to \infty$), viz.,
\begin{equation} 
\lim_{h \rightarrow 0} J = \left\{\begin{array}{ll}
A\kappa r_0^{d/2}\Big(\frac{r_0^{(4-d)/2}}{u_0}\Big)^{d/2}\rho^{d(d-3)/2}
\exp{\Big[-\frac{r_0^{(4-d)/2}}{u_0}\big(c\rho^{d-1}+\dots\big)\Big]}
& d \ne 3 \\ \\ A\kappa r_0^{3/2}\Big(\frac{r_0^{1/2}}{u_0}\Big)^{3/2}\rho^{7/3}
\exp{\Big[-\frac{r_0^{1/2}}{u_0}\big(c\rho^{2}+\dots\big)\Big]} & d=3
\end{array}\right.
\label{eq:gnw-last}
\end{equation}
\noindent
Now by transcribing the $\psi^4$-theory parameters to the pertinent
variables in alloys, we identify $\alpha_d^{1/(d-1)}\rho
\Leftrightarrow (x_0/x)$, where $\alpha_d \equiv r_0^{(4-d)/2}/u_0$ is
an auxiliary variable and the variables $x$ and $x_0$ are defined in
section \ref{sec:nuke}. Setting $\xi = r_0^{-1/2}$ (with $K=2$), we
convert Eq. (\ref{eq:gnw-last}) to
\begin{equation} 
J \sim \left\{\begin{array}{ll}
 \alpha_d^{d/(d-1)}\kappa\xi^{-d}\Big(\frac{x_0}{x}\Big)^{d(d-3)/2}
\exp{\Big[-\Big(\frac{x_0}{x}\Big)^{d-1}\Big]}
& d \ne 3 \\ \\ \alpha_3^{1/3} \kappa \xi^{-3}\Big(\frac{x_0}{x}\Big)^{7/3}
\exp{\Big[-\Big(\frac{x_0}{x}\Big)^{2}\Big]} & d=3
\end{array}\right.
\label{eq:gnw-trans}
\end{equation}
\noindent
From here on, to obtain Eq. (\ref{eq:nr_gnw}) is
fairly straightforward, as it has been explicitly shown by Langer and
Schwartz \cite{Langer_Schwartz_1980}.
\section{Nucleation Delay Time}
\label{sec:appendc}
In order to calculate the time-lag for nucleation in classical theory,
one needs to solve the full Fokker-Planck equation (\ref{eq:FPE}). Here,
however, we can make a rough estimate of this quantity for the case of
$t>>B^{-1}$, meaning that; we are interested in time intervals much
larger than the time of relaxation ($ \propto B^{-1}$). We
assume that $n=n(R,t)$ obeys:
\begin{equation}
\frac{\partial n}{\partial t} =  B(R_c)\frac{\partial^2 n}{\partial R^2}\,,
\label{eq:dife}
\end{equation}
\noindent
subject to the boundary conditions $n(0,t)= n_1$, $n(s,t) = 0$ and the
initial condition $n(R,0) = 0$ for $s \ge R_c$.
The solution of this boundary value problem is expressed as
\begin{equation}
  \label{eq:sol_1}
  n(R,t) =
  n_1\big(1-\frac{R}{s}\big)-\frac{2n_1}{\pi}\sum_{m=1}^\infty
  \frac{1}{m}\sin{\Big(\frac{m\pi R}{s}\Big)}e^{-\frac{m^2\pi^2 B_c t}{s^2}}.
\end{equation}
\noindent
The nucleation rate is related to the gradient of $n(R,t)$, viz.,
\begin{equation}
\label{eq:jrc_1}
  J(s,t) \approx  -B_c\frac{\partial n}{\partial R}\bigg\arrowvert_{R=s}
  =\frac{B_cn_1}{s}\vartheta_4(0,q),
\end{equation}
\noindent
where $\vartheta_4(u,q)=1+2\sum_{m=1}^{\infty}(-1)^m q^{m^2}\cos{(2mu)}$ is an
elliptic theta function \cite{Whittaker_Watson_1927}, here with
$q\equiv\exp{(-B_c\pi^2 t/s^2)}$. Employing the Courant-Hilbert identity
\cite{Courant_Hilbert_1966},
\begin{equation}
\label{eq:ch-id}
\sum_{m=-\infty}^{\infty}\exp{(-\pi m^2 \zeta)}(-1)^m =
\frac{1}{\sqrt{\zeta}}\sum_{m=-\infty}^{\infty}\exp{[-\frac{\pi
    (m-1/2)^2}{\zeta}]},
\end{equation}
\noindent
Eq. (\ref{eq:jrc_1}) can be expressed in the form
\begin{equation}
\label{eq:jrc_3}
  J(s,t)=\frac{B_c n_1}{\sqrt{\pi}s}(\frac{t_d}{t})^{1/2}
\,\vartheta_2(0,e^{-\frac{t_d}{t}}),
\end{equation}
\noindent
where $t_d\equiv s^2/B_c$ is the delay time. For short
times, $t<<t_d$, power series expansion of the theta function gives
\begin{equation}
\label{eq:jrc_4}
  J(s,t)\sim\frac{2B_cn_1}{\sqrt{\pi}s}\big(\frac{t_d}{t}\big)^{1/2}
\big[e^{-\frac{t_d}{t}}+\mathcal{O}(e^{-\frac{9\,t_d}{4t}})\big].
\end{equation}
Equation (\ref{eq:jrc_3}) offers a zeroth approximation to the
time-dependence of the nucleation rate, however, it illustrates the
essential physics of the phenomenon.  Detailed calculations of the
time-dependent nucleation rate and the delay time, within the
classical nucleation theory, can be found in Refs.
\cite{Binder_Stauffer_1976, Kashchiev_1969}.



\bibliography{alma}

\end{document}